# Human-in-the-Loop Design Cycles – A Process Framework that Integrates Design Sprints, Agile Processes, and Machine Learning with Humans

Chaehan So[1]

[1]Information & Interaction Design,
Humanities, Arts & Social Sciences Division,
Yonsei University, 03722 Seoul, South Korea
`cso@yonsei.ac.kr`

**Abstract.** Demands on more transparency of the backbox nature of machine learning models have led to the recent rise of *human-in-the-loop* in machine learning, i.e. processes that integrate humans in the training and application of machine learning models. The present work argues that this process requirement does not represent an obstacle but an opportunity to optimize the design process. Hence, this work proposes a new process framework, Human-in-the-learning-loop (HILL) Design Cycles – a design process that integrates the structural elements of agile and design thinking process, and controls the training of a machine learning model by the human in the loop.
The HILL Design Cycles process replaces the qualitative user testing by a quantitative psychometric measurement instrument for design perception. The generated user feedback serves to train a machine learning model and to instruct the subsequent design cycle along four design dimensions (novelty, energy, simplicity, tool).
Mapping the four-dimensional user feedback into user stories and priorities, the design sprint thus transforms the user feedback directly into the implementation process. The human in the loop is a quality engineer who scrutinizes the collected user feedback to prevents invalid data to enter machine learning model training.

## 1 Introduction

Design Thinking has established itself as a proven process to create innovative products from the end user perspective [1], commonly denoted *as user-centered design* [2] or *human-centered design* [3]. Despite the uncontested merits of design thinking for innovation [4], many online companies do not follow this methodology yet. In their view, it is incompatible with their product development process or machine learning system. The present work aims to propose a process framework for such companies that integrates design thinking with a development process that incorporates the human in the loop.

Why has it become essential in 2020 to re-integrate the human in the loop in machine learning systems? The underlying reason is that in recent years, the explosive growth of AI research results also led to increasing doubts on the validity of these findings. Apart from the concerns on *reproducibility* and *interpretability* [5], a new concern is on *controllability*, i.e. the demand for integrating humans in any process that relies on the outcome of machine learning, now referred to as *humans-in-the-loop* [6].

Two of the most crucial questions, about how humans in the loop should affect machine learning conceptually, were asked in a workshop on human-centered machine learning at the CHI 2016 conference by Gillies and colleagues [7]:

1. What is the role of humans in existing machine learning systems?
2. How does a human-centered approach change the way machine learning is done?

The present work suggests as an answer to the preceding questions to shift part of the learning process from machine learning to conventional psychological methodology. This shift allows not only for controlling the machine learning process but also speed up the human learning loop by more active involvement.

The proposed framework applies psychometrics to understand users' design perception to generate feedback for the subsequent design cycle, and to incrementally update a machine learning model to accommodate the fast-changing nature of user preferences in the online world. These elements are enabled by merging design thinking methodology with an agile process.

The inherent reason why online companies cannot implement design thinking methodology may lie in misunderstandings of its process. For example, the Stanford d.school design thinking process, introduced by IDEO in 2007 [8], specifies a cycle of five process phases from emphasize, define, ideate, prototype, and test. The first misunderstanding of many companies is to disregard the cyclical nature and instead implement a one-time execution of the five phases. The final test phase thus does not feed a subsequent learning cycle but merely serves as an end-to-end system test. Nevertheless, the cyclical nature can easily be implemented by an *agile process* [9], i.e. an iterative and incremental development process. The second misunderstanding occurs when businesses do not implement design thinking as an iterative learning cycle. Design thinking methods usually specify the sample size of a qualitative user test between five to ten people. As the testing phase is only performed once, the small sample size results in feedback that is not representative and thus risky to base product decisions on. The user feedback obtained in such a small qualitative workshop is prone to positive distortion as the participants tend to give feedback that pleases the makers of the presented prototypes. In conclusion, innovative companies have an understandable need for representative user feedback that is informative, i.e. provide directions for the implementation process.

Taken together, the preceding considerations lead to the following research question:

> *How can humans in the loop integrate into an agile process*
> *with design thinking and machine learning methodology?*

## 2    Building Blocks

### 2.1    Psychometrics

In psychological studies, people's behaviors and perceptions are analyzed by qualitative [10] and quantitative research methodology [11]. The qualitative methodology is largely overlapping with design research [12], e.g. anthropological methods of investigating user contexts, or qualitative structured and semi-structured user interviews [10]. One major disadvantage of qualitative methods is their vulnerability to various cognitive and emotional bias effects. One such frequently encountered bias is *confirmation bias* [13], i.e. the tendency to ignore facts contradictory to prior judgment. Researchers exert confirmation bias e.g. when they ask *leading questions* [14].

Quantitative psychological research methodology aimed to reduce biases by a range of statistical tools summarized under the term *psychometrics* [15], the science of psychological measurement. One of its focus areas is the analysis of surveys from self-reported user perceptions [16]. E.g. it specifies how to find in survey responses an underlying pattern of psychological factors by *explorative factor analysis* [17].

### 2.2    Psychometric Measurement Instrument for Design Perception

The proposed framework is based on the author's prior work [8] that developed a measurement instrument for design perception from 1955 design works evaluations. It consists of 12 items in four design dimensions extracted by principal factor analysis.

The development of this measurement instrument followed psychometric scale construction methodology. This entailed that the initial item pool, generated by qualitative methodology, was systematically reduced by analyzing the factor structure by explorative factor analysis. The latter method investigates the items (survey questions) to detect statistical patterns that could correspond to underlying psychological factors [11]. The items displayed a so-called *simple structure* [15], i.e. the factor structure revealed that all items show high *convergent validity* (items for one construct load on the same component) as well as *discriminant validity* (items for different constructs load on different components).

For each of the four factors, a plausible interpretation could be found (*novelty*, *energy*, *simplicity*, and *tool*). Furthermore, when analyzed as a separate scale, each of these factors showed high reliability according to the convention by Cohen [18] with Cronbach alpha coefficients around 0.80. Together, these factors explain most of the variance in the survey responses (74%). The items are displayed in Table 1.

**Table 1.** Design Dimensions – Items

| Novelty | Energy | Simplicity | Tool |
|---|---|---|---|
| *exciting* | *powerful* | *simple* | *practical* |
| *unique* | *clever* | *clear* | *functional* |
| *creative* | *intuitive* | *minimalistic* | *useful* |

### 2.3 The Human in the Loop

Nushi and colleagues from ETH Zurich and Microsoft Research [19] disentangle machine learning systems into components that allow their troubleshooting methodology to locate the component responsible for failure to enable targeted fixes of that component. Their framework incorporates humans in the loop to simulate component fixes and evaluate the machine learning system before and after the fixes to reach a higher level of control. According to the authors, the preconditions for such a concept are the modularity of a machine learning system and the interpretability of its components.

Xin and colleagues [20] from University of Illinois, Urbana-Champaign, presented a concept for tackling the "tedious process of iterative experimentation" at a workshop at the Data Management for End-To-End Machine Learning (DEEM) 2018 conference. They point to several inefficient aspects of current machine learning systems that do not cater to human-in-the-loop in the iterative development process. Their framework caters to the developers' need for reusing intermediate results as opposed to rerunning the whole workflow. In particular, developers need to understand the impact of their code changes early as opposed to waiting until the computation is fully finished. Taken together, if a machine learning system integrates its developers, the humans in the loop, better by providing them with "rapid, approximate feedback", it will result in a significant speedup of the end-to-end process.

Taken together, machine learning should only be trained if humans in the loop have verified the validity of the input data, and learned from the model training by early approximate feedback. The approximate feedback generation may also allow for discarding older input data which may have become obsolete.

### 2.4 Agile Process

An agile process is mainly characterized by an iterative and incremental approach to development [21, 22]. This means that the product development process is not defined by a series of long, subsequent phases like in the waterfall model, but as a series of short temporal cycles called iterations [23]. Scrum, the most common agile method, calls these iterations *sprints* [24]. Google developed its design thinking method based on agile processes and thus called it *design sprint* [25].

One essential process aspect, *time-boxing* [26], is frequently misunderstood and thus incorrectly implemented. *Time-boxing* [27] specifies that the iteration duration is strictly fixed. In the real world, this means that the implementation team is not allowed to delay the iteration finish date despite the pressuring demands by project managers and business teams to obtain a planned scope at the end of an iteration. Instead of fixing the scope to be delivered, time-boxing allows for adjusting the scope to keep the timeline.

Scrum and the Google design sprint describe the iteration planning in a workshop called *sprint planning*. In this planning workshop, the implementation team defines the scope of the current sprint based on the customer feedback for the last sprint results. The new scope is fleshed out into *user stories*, i.e. requirements formulated from the user perspective [28]. All user stories are prioritized according to customer feedback.

## 3   Proposed Process Framework

This work proposes a process framework, as depicted in Figure 1, that consists of a design thinking process, merged into an agile development process, that replaces the qualitative user test by quantitative measurement of user feedback. This replacement allows to generate scalable and instructional feedback for the subsequent learning cycle which is implemented as a design sprint.

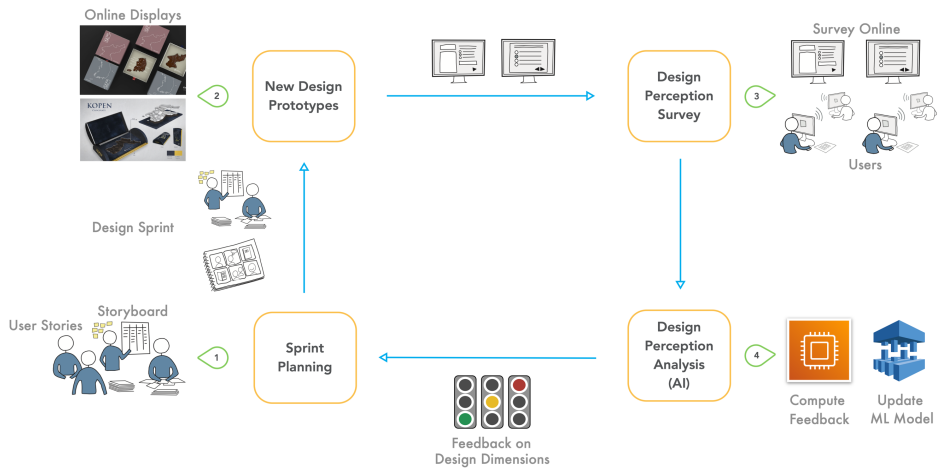

**Fig. 1**. Process Framework: Human-in-the-Loop Integration of End Users into a Machine Learning-Based Analysis Process

### 3.1   User Testing – by Design Perception Survey

For the user testing, the team acquires a pool of the company's real end users, and invites a portion of this user pool at the design sprint end to answer an online survey. This online survey displays the new prototypes which were produced in the preceding design sprint (see Figure 2). The invited survey participants asses the 12 items of the design perception measurement instrument [29] about the displayed new prototypes. Additionally, the survey solicits qualitative feedback such as new feature requests or questions about functionality details.

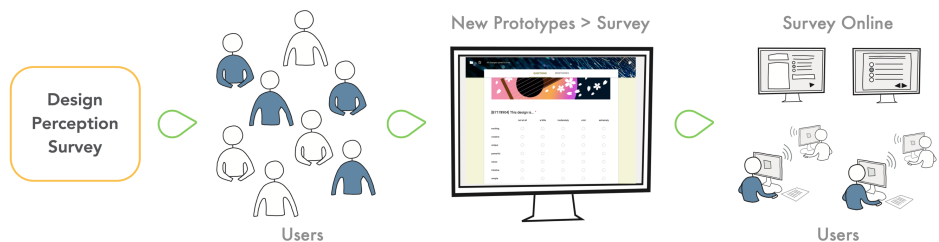

**Fig. 2.** User Testing by Design Perception Survey

### 3.2 Feedback Computation and Machine Learning Model Update – by Design Perception Analysis

The responses to the design perception survey are analyzed in the following manner. The users' scale responses are grouped according to the four design dimensions, e.g. exciting, unique, creative to novelty. For each design dimension scale, the composite scores are calculated. These composite scores represent the scale scores from each user and can be displayed by boxplots which provide visualized feedback to the implementation team (Figure 3).

The human-in-the-loop aspect is realized by a human quality engineer who scrutinizes the received user responses in data quality, and discards invalid data like outliers or responses containing strong acquiescence bias. This data cleaning procedure is essential to retain only valid new data to add to the dataset on which the machine learning model is trained. The resulting model can serve as a pretrained model for quick simulations to support prototyping decisions.

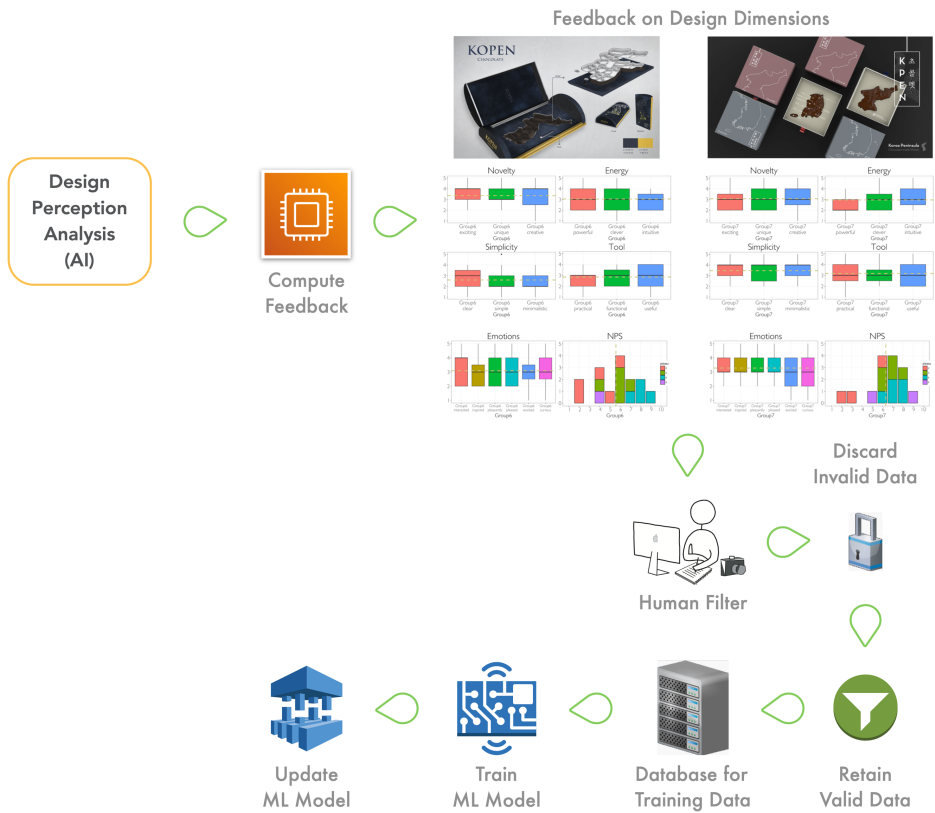

**Fig. 3.** Design Dimensions Feedback Generation & Machine Learning Model Update fed by Design Perception Analysis

### 3.3 Sprint Planning – by Design Dimensions Feedback

The sprint planning process is structured by the four design dimensions such that each design dimension corresponds to a category in the sprint storyboard (see Figure 4).

The category (e.g. simplicity) refers to a high-level abstraction of scope or business requirements respectively. The team assigns the priority of each category according to the design dimensions' composite score in the computed feedback – the lowest score gives the highest priority because it uncovers the strongest deficiency.

In descending priority of the design dimensions, the team decides which design dimensions shall be addressed in the upcoming sprint. For this decision, the team does not have to consider the ease or difficulty of implementation because the latter is reflected in the later effort estimation process. This means that user stories that are easier to implement will get less effort estimation units and will consequently be more likely implemented.

The team writes user stories for the selected design dimensions. For example, a user story in category simplicity could be formulated as "*As a frontend web user, I want to navigate to my personal page with the least possible number of navigational steps*".

When writing such user stories, the team integrates the qualitative user feedback into the user stories' *acceptance criteria*. For example, if user feedback hinted to inconsistent colors, an acceptance criterion for a user story in category simplicity could be formulated as *"Check if all UI elements originate from the same color scheme"*.

Based on the user stories, the team conducts the *agile effort estimation process* [30] and adjusts the sprint scope accordingly. It then performs the *task breakdown*, i.e. team members break down the scope defined in the user stories into small tasks that must be performed to implement the user story. After the task breakdown, it reviews the team's understanding of the current sprint scope and concludes the sprint planning session.

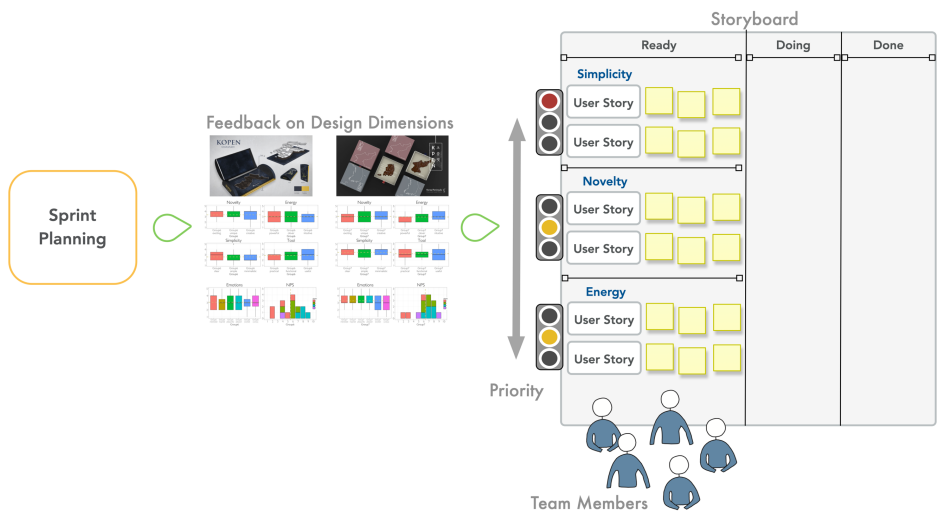

**Fig. 4.** Sprint Planning by Design Dimension Feedback

### 3.4 New Design Prototypes – by Design Sprints

The team executes the design sprint as a normal agile iteration as a self-organized team. At the end of this process (see Figure 5), the team applies special care to present its results in the subsequent survey because the users must be able to grasp the new functionality and from the online display. This care entails additional activities, like taking photos or producing renderings from different view angles, until users can grasp the gist of the new prototypes in their online survey displays (see Figure 5, prototype displayed on the bottom middle in Figure 5).

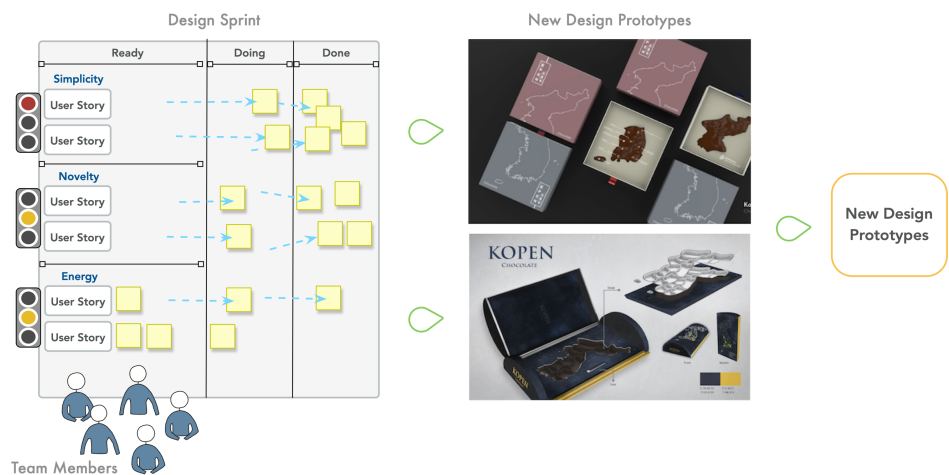

**Fig. 5.** New Design Prototypes by Design Sprints

## 4 Discussion

Many companies of online products or services that rely on a machine learning system do not know how to integrate humans in the loop and often assume that this would decrease efficiency. Likewise, companies that follow an agile product development process have difficulty integrating this process with design thinking methodology. The present work provides a solution for both these practical problems by proposing a new process framework, the Humans-in-the-Learning-Loop (HILL) Design Cycles.
Several of its process elements go hand in hand to create synergic effects:

1. The design prototypes are used in online surveys. This enables high scalability of the user testing despite the collection of qualitative feedback by the same survey. Moreover, it allows collecting other user data which are used for the machine learning training.
2. The user test results are analyzed with a psychometric measurement instrument that generates quantitative feedback. The quantitative nature allows to directly use it in machine learning model training. Furthermore, this feedback is on a summarized, structured level that is informative for the next design sprint because it provides directions by clear scores on design dimensions.

3. The quality engineer who reviews the generated feedback caters both for data quality improvement as well as for early detection of trends, e.g. changing user preferences.
4. The machine learning model is updated in each design sprint. Therefore, decision makers can always rely on the most recent user data and do not have to wait for the completion of a long training period [20].

The HILL Design Cycles merge the design thinking method into an agile process, and leverage the user feedback with the human in the loop for machine learning model training and directions for the design sprint.

**Acknowledgment**

This research was supported by the Yonsei University Faculty Research Fund of 2019-22-0199.